\documentclass[journal]{IEEEtran}
\pdfoutput=1
\usepackage [latin1]{inputenc}
\usepackage{subfigure}
\usepackage{bm}
\usepackage{graphicx}
\usepackage{amsmath, amsthm, amsfonts, amssymb, amsbsy}
\usepackage{setspace}
\usepackage{graphicx}
\usepackage{pstricks,arydshln}
\usepackage{multirow}
\usepackage{booktabs}
\usepackage{stfloats}
\usepackage{mathbbol,float,algorithmic,amscd,textcomp,amssymb,epsfig,graphics,epsf,color,amsmath,balance,algorithm,cite}
\usepackage{caption}
\usepackage{cite}
\usepackage{xcolor}
\usepackage{epsfig}
\usepackage{epstopdf}
\usepackage{gensymb}
\usepackage{multirow}
\usepackage{booktabs} 

\newtheorem{rem}{Remark}

\newtheorem{thm}{Theorem}

\begin{document}

\title{Broadband Beam Steering for Misaligned Multi-Mode OAM Communication Systems (Special Section on SASP)}%

\author{Zhengjuan Tian,~\IEEEmembership{Student Member,~IEEE,}~Rui Chen,~\IEEEmembership{Member,~IEEE,}~Wen-Xuan Long~\IEEEmembership{Student Member,~IEEE,}\\
Hong Zhou~\IEEEmembership{Student Member,~IEEE} and Marco Moretti,~\IEEEmembership{Member,~IEEE}
\thanks{R. Chen is with the State Key Laboratory of ISN, Xidian University, Xi'an 710071,
China, and also with the National Mobile Communications Research Laboratory,
Southeast University, Nanjing 210018, China (e-mail: rchen@xidian.edu.cn).}
\thanks{Z. Tian, W.-X. Long and H. Zhou are with the State Key Laboratory of ISN, Xidian University, Shaanxi 710071, China (e-mail: \{zjtian,wxlong,hzhou\_1\}@stu.xidian.edu.cn).}
\thanks{M. Moretti is with the University of Pisa, Dipartimento di Ingegneria dell'Informazione, Italy (e-mail: marco.moretti@iet.unipi.it).}
}

\maketitle

\begin{abstract}
Orbital angular momentum (OAM) at radio frequency (RF) has attracted more and more attention as a novel approach of multiplexing a set of orthogonal OAM modes on the same frequency channel to achieve high spectral efficiency (SE). However, the precondition for maintaining the orthogonality among different OAM modes is perfect alignment of the transmit and receive uniform circular arrays (UCAs), which is difficult to be satisfied in practical wireless communication scenario. Therefore, to achieve available multi-mode OAM broadband wireless communication, we first investigate the effect of oblique angles on the transmission performance of the multi-mode OAM broadband system in the non-parallel misalignment case. Then, we compare the UCA-based RF analog and baseband digital transceiver structures and corresponding beam steering schemes. Mathematical analysis and numerical simulations validate that the SE of the misaligned multi-mode OAM broadband system is quite low, while analog and digital beam steering both can significantly improve the SE of the system. However, digital beam steering can obtain higher SE than analog beam steering especially when the bandwidth and the number of array elements are large, which validates that baseband digital transceiver with digital beam steering is more suitable for multi-mode OAM broadband wireless communication systems in practice.
\end{abstract}

\begin{IEEEkeywords}
Orbital angular momentum (OAM), uniform circular array (UCA), broadband wireless communication, misalignment, beam steering
\end{IEEEkeywords}

\section{Introduction}
Currently, explosive growth of emerging services, e.g., high-definition (HD) video, virtual reality (VR), auto-pilot driving and Internet of Things (IoT), calls for a large increase of wireless data rate. However, it is inadequate to utilize the traditional techniques to catch up with the growing requirement of the data rate. Therefore, a number of effective techniques are developed.
One available technique, called spectrum extension, has attracted tremendous attention in recent years, for instance, millimeter wave and terahertz bands are being licensed \cite{WRC,Song2011Present}. It is well-known that the initial commercial deployments of fifth generation (5G) New Radio (NR) networks, both at sub-6 GHz and at mmWave frequencies, are already under way during 2019 \cite{Ghosh20195G}, which will adopt larger bandwidth (e.g., 100MHz for sub-6GHz band, and 400MHz for mmWave band) than forth generation (4G) networks to support three generic types of connectivity: extended mobile broadband (eMBB), ultra-reliable low-latency communication (URLLC) and massive machine-type communication (mMTC) \cite{Popovski5G,3GPP2017}. Since the radio frequency spectrum resources are scarce, besides exploiting more frequency bandwidth, some innovative techniques, such as advanced modulation and coding schemes, intelligent cognitive radio (CR) and massive multiple-input multiple-output (MIMO), have been explored to enhance the system spectral efficiency (SE).

In essence, almost all existing wireless communication techniques are based on planar electromagnetic (EM) waves. Since the discovery in 1992 that light beams with helical phase fronts can carry orbital angular momentum (OAM) \cite{Allen1992Orbital}, a significant research effort has been focused on vortex EM waves as a novel approach for multiplexing a set of orthogonal OAM modes on the same frequency channel and achieving high SE {\cite{Tamburini2012Encoding,Mahmouli20134,Yan2014High,Zhang2016The,Ren2017Line, Zhang2017Mode,Chen2018A,Chen2018Beam,Chen2019On,Zhang2019Orbital,Chen2019Orbital,Zhao2019Compound,Chen2020OAM,Chen2020Multi,Chen2020Generation}. The OAM beams can be generated by spiral phase plate, spiral parabolic antenna, metasurface material and uniform circular array (UCA) \cite{Chen2019Orbital}, where UCA is more common due to the flexibility of generating and receiving multi-mode OAM beams and realizing beam steering. Specifically, the single-mode OAM beam can be generated by a UCA through feeding its antenna elements with the same input signal, but with a successive phase shift from element to element \cite{Edfors2012Is}. On this basis, after a full turn the phase has the increment of $2\pi\ell$, where $\ell$ is an unbounded integer termed as OAM mode number \cite{Allen1992Orbital}. Thus, the multi-mode OAM beam can be achieved by synthesizing multiple single-mode beams through RF analog or baseband digital approaches with corresponding transceiver structures \cite{Chen2018A}.

In spite of vortex EM waves carrying OAM being deemed as a promising Beyond 5G (B5G) technique \cite{Zhang20196G,Yang20196G}, there are still some technical challenges for the practical application of OAM wireless communication.
One challenge is that OAM wireless communication requires perfect alignment between the transmit and receive antenna arrays. It is analyzed in \cite{YjZhang2013,Xie2015Performance,Chen2018Beam} that for OAM narrow-band communication systems, if the precondition is not accurately met, the system performance quickly deteriorates, and beam steering method proposed in \cite{Chen2018Beam} can avoid the performance deterioration caused by misalignment. However, the effect of misalignment on the performance of OAM broadband systems and corresponding beam steering method have not been studied. As a widely accepted use case of the OAM communication, future extremely high data-rate backhaul transmission will definitely adopt large bandwidth.

Therefore, in this paper, we first present the OAM broadband communication system models for the UCA-based RF analog and baseband digital transceiver structures in a more general non-parallel misalignment scenario. Then, we analyze the effect of the oblique angles on the performance of multi-mode OAM broadband communication system from the perspective of channel gain and inter-mode interference (IMI). Thereafter, we compare two beam steering schemes, i.e. analog beam steering (ABS) and digital beam steering (DBS), corresponding to the two transceiver structures, in mitigating the destructive effect of oblique angles. At last, we validate that both ABS and DBS can circumvent large performance deterioration, but baseband digital OAM transceiver with DBS has better performance for the misaligned multi-mode OAM broadband communication system.

{\sl Notations}: Upper (lower) case boldface letters are used to denote matrices (vectors), non-bold letters are used to denote scalar values. $(\cdot)^T$ and $(\cdot)^H$ denote transpose and conjugate transpose, respectively. $\otimes$ and $\odot$ denote the Kronecker product and Hadamard product, respectively. $\mathbf{I}_n$ is a $n\times n$ identity matrix and $\mathbf{1}_n$ is a $n\times 1$ vector of ones. $|\cdot|$ denotes modulus. $\mathbb{E}\{\cdot\}$ and $\mathbb{Z}$ represent expectation and integer field, respectively.
\begin{figure}[t]
\centering
\subfigure[]
{\includegraphics[width=8.5cm]{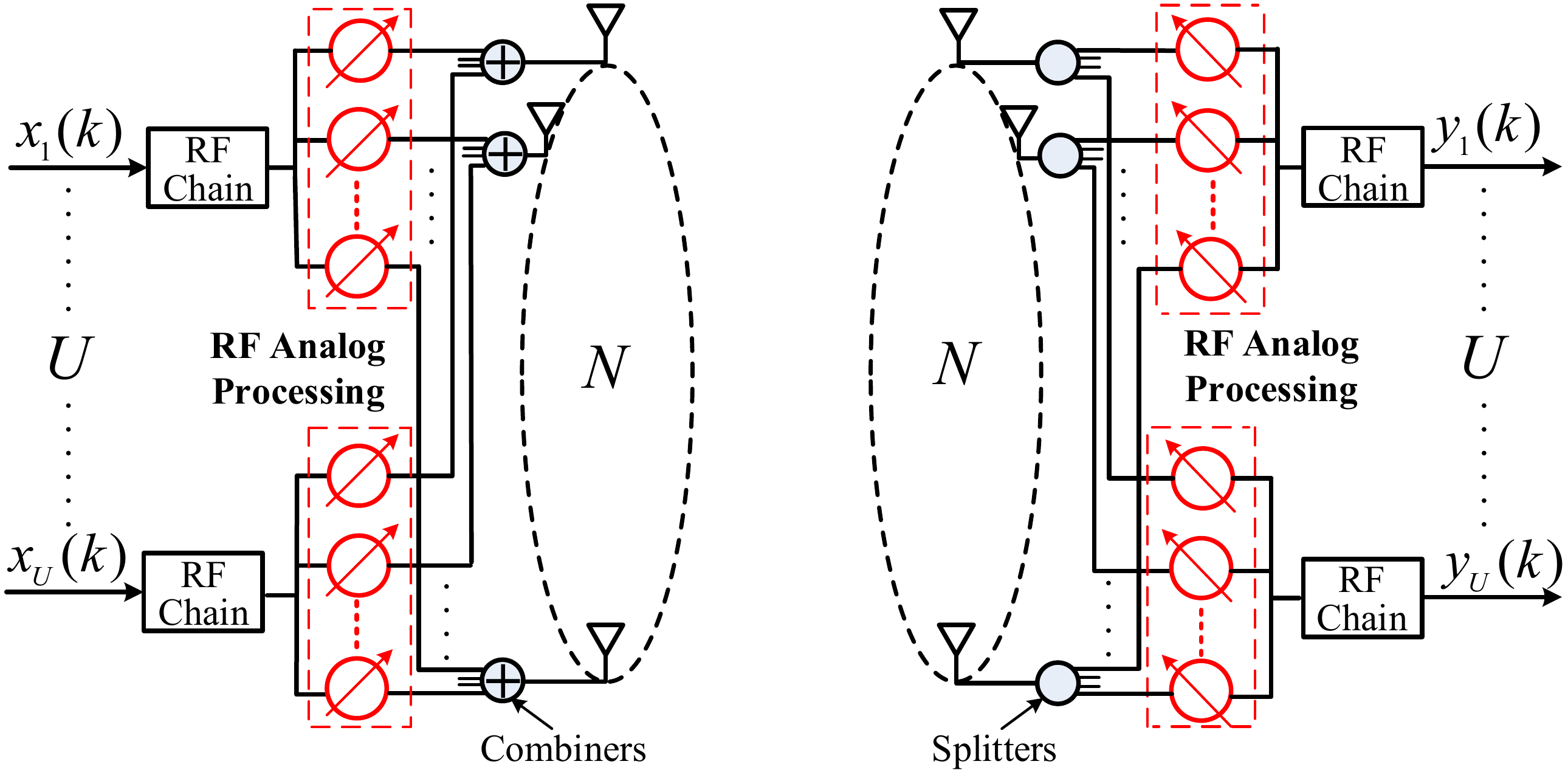}}

\subfigure[]
{\includegraphics[width=8.5cm]{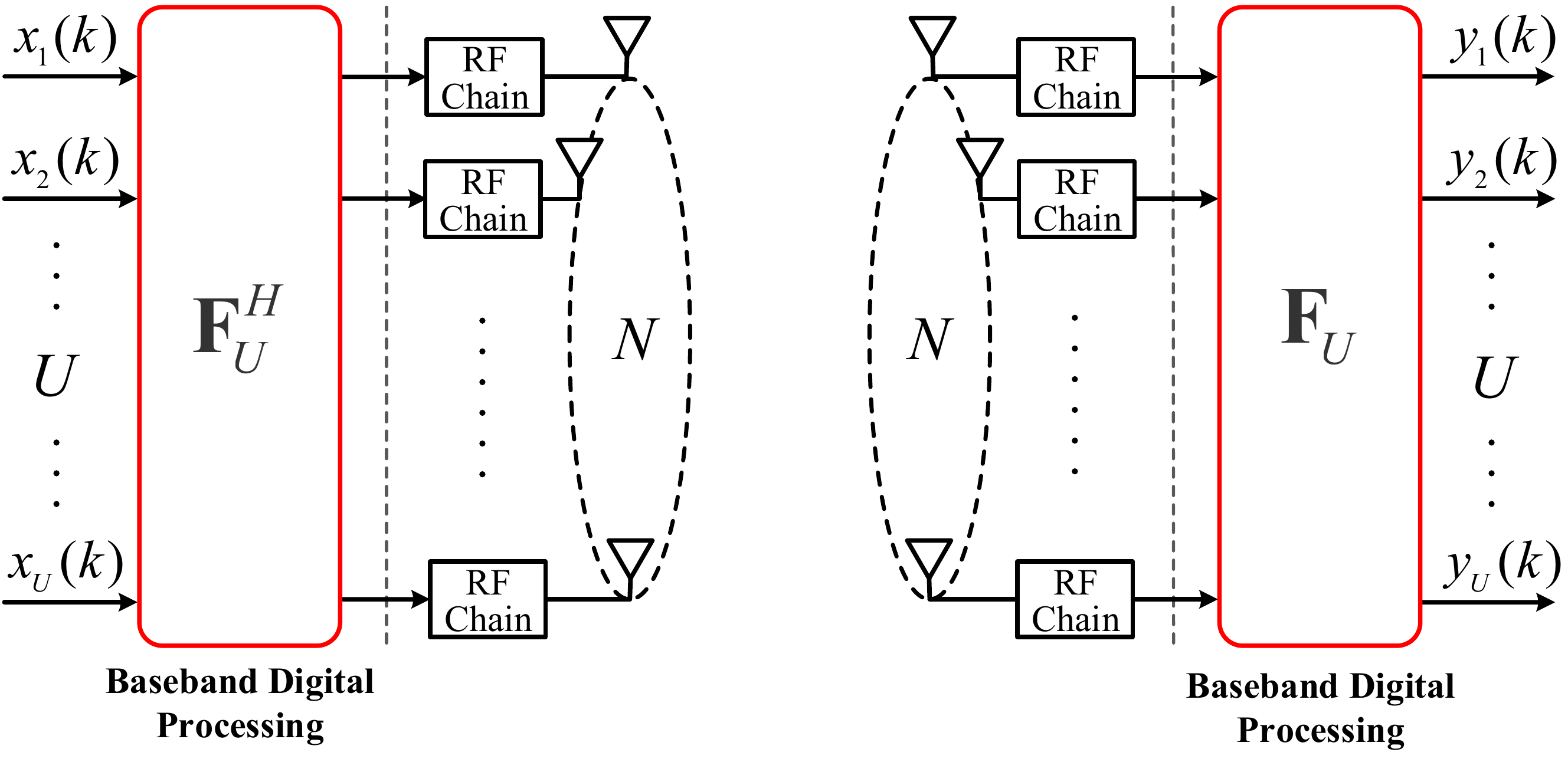}}
\caption{Simplified block diagram of UCA-based multi-mode OAM transceiver implemented by (a) RF analog processing, and (b) baseband digital processing. }
\label{Fig1}
\end{figure}

\section{System Model}
We consider a multi-mode OAM broadband communication system, where the multi-mode OAM beam is generated by an $N$-elements UCA at the transmitter and received by another $N$-elements UCA at the receiver.

\subsection{Multi-Mode OAM Broadband OFDM Systems}
\label{BCS}
It is well known that orthogonal frequency division multiplexing (OFDM) technology is widely used in 4G networks, and is also the core technology of 5G NR, which has been specified in the physical layer fundamentals of 5G NR non-standalone air interface. In order to address diverse scenarios and deployments, 5G NR will adopt a scalable OFDM technology, where the OFDM waveform is scalable in the sense that the subcarrier spacing of OFDM can be chosen according to $15 \times 2^n$kHz, $n$ is a specified integer and 15kHz is the subcarrier spacing used in Long Term Evolution (LTE) \cite{Zaidi2018OFDM}. Therefore, we consider OFDM technology for our OAM broadband communication system to support the generic types of connectivity in 5G NR.

In practice, the perfect alignment between the transmit UCA and the receive UCA are difficult to realize. For an arbitrary misalignment case, the beam steering needs performing at both the transmitter and the receiver \cite{Chen2018Beam}. Corresponding to the two transceiver structures shown in Fig. \ref{Fig1}, there are two beam steering schemes, i.e., ABS and DBS, which are implemented by RF analog transceiver structure and baseband digital transceiver structure, respectively.
It's apparent that the phase shifts induced by the phase shifters of the RF analog transceiver are fixed for the broadband signals to be transmitted at different frequencies, while for the baseband digital transceiver, phase shifts are easy to be applied to the broadband OFDM signals at different subcarriers. Hence, the beam steering matrices of ABS at the transmitter and receiver are denoted as $\mathbf{W}_A$ and $\mathbf{P}_A$, while the beam steering matrices of DBS at the transmitter and receiver denoted as $\mathbf{W}_D(k_p)$ and $\mathbf{P}_D(k_p)$ are functions of the subcarrier frequency, where $k_p=2\pi/\lambda_p$ is the wave number at the $p$th subcarrier, and $\lambda_p$ is the wavelength at the $p$th subcarrier.

Thus, in the transmission of $U$-mode OAM beam that multiplexes $U$ OAM modes in the free space channel at the $p$th subcarrier, the despiralized OFDM symbol vectors at the RF analog and baseband digital receivers, denoted as $\mathbf{y}_A(k_p)$ and $\mathbf{y}_D(k_p)$ respectively, take the form
\begin{equation} \label{ya}
\mathbf{y}_A(k_p)= \mathbf{W}_A\odot\mathbf{F}_U \left(\mathbf{H}(k_p)\mathbf{F}_U^{H}\odot \mathbf{P}_A\mathbf{x}(k_p)+\mathbf{z}(k_p) \right),
\end{equation}
\begin{align} \label{yd}
\mathbf{y}_D(k_p) = &\mathbf{W}_D(k_p)\odot\mathbf{F}_U \big(\mathbf{H}(k_p)\mathbf{F}_U^{H}\odot\mathbf{P}_D(k_p)\mathbf{x}(k_p) \nonumber\\
&+\mathbf{z}(k_p) \big),
\end{align}
where $\mathbf{y}_A(k_p)\!=\![y_A(k_p,\ell_1), y_A(k_p,\ell_2), \cdots, y_A(k_p,\ell_U)]^T$, $\mathbf{y}_D(k_p)\!=\![y_D(k_p,\ell_1), y_D(k_p,\ell_2), \cdots, y_D(k_p,\ell_U)]^T$, $\mathbf{x}(k_p)=[x(k_p,\ell_1), x(k_p,\ell_2), \cdots, x(k_p,\ell_U)]^T$, $x(k_p,\ell_u)$ is the baseband modulation symbol transmitted on the $u$th mode OAM beam at the $p$th subcarrier, $y_A(k_p,\ell_u)$ and $y_D(k_p,\ell_u)$ are respectively the received modulation symbols at the $u$th mode and $p$th subcarrier at the receiver, $\mathbf{F}_U=[\mathbf{f}^H(\ell_1), \mathbf{f}^H(\ell_2),$ $\cdots,\mathbf{f}^H(\ell_U)]^H$ is a $U\times N$ partial Fourier matrix, $\mathbf{f}(\ell_u)=\frac{1}{\sqrt{N}}[1,e^{\frac{-j2\pi \ell_u}{N}},\cdots,e^{\frac{-j2\pi \ell_u (N-1)}{N}}]$, $\mathbf{z}(k_p)=[z(k_p,1),z(k_p,2),\cdots,z(k_p,N)]^T$ is the complex Gaussian noise vector with zero mean and covariance matrix $\sigma^2_z\mathbf{I}_{N}$, $p=1,2,\cdots,P$, $u=1,2,\cdots,U$, and $U\leq N$.


For easier analysis, we only consider beam steering at the receiver as shown in Fig. \ref{Fig2} here, and the arbitrary misalignment case could be considered and verified similar to that in \cite{Chen2018Beam}. Hence, according to \eqref{ya} and \eqref{yd}, the despiralized OFDM symbol vectors at the RF analog and baseband digital receivers can be written as
\begin{equation} \label{ya1}
\mathbf{y}_A(k_p)=\mathbf{W}_A \odot \mathbf{F}_U \left(\mathbf{H}(k_p)\mathbf{F}^H_U \mathbf{x}(k_p)+\mathbf{z}(k_p) \right),
\end{equation}
\begin{equation} \label{yd1}
\mathbf{y}_D(k_p)= \mathbf{W}_D(k_p)\odot \mathbf{F}_U \left(\mathbf{H}(k_p)\mathbf{F}^H_U \mathbf{x}(k_p)+\mathbf{z}(k_p)\right).
\end{equation}
\begin{figure}[t]
\begin{center}
\includegraphics[width=7.5cm]{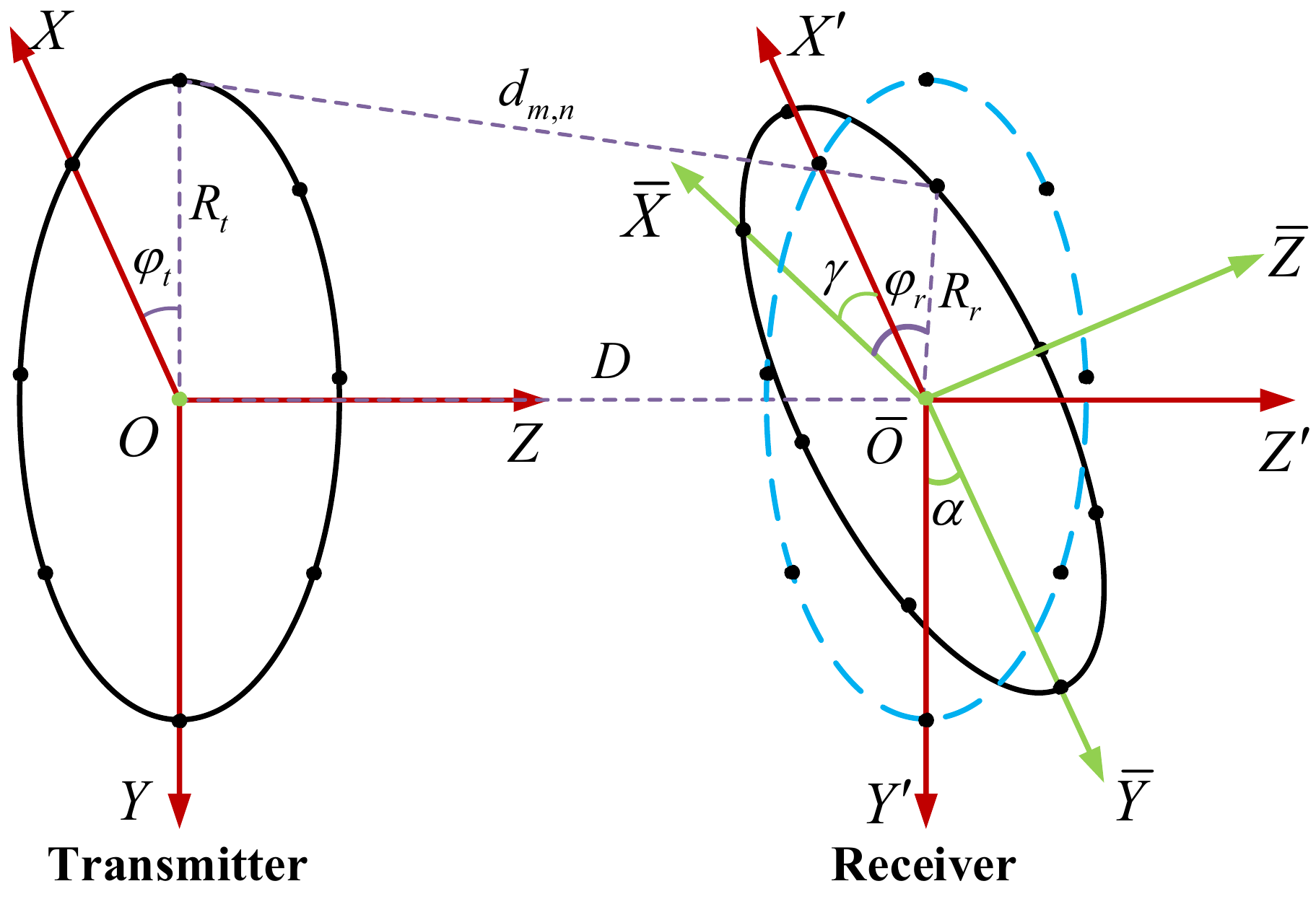}
\end{center}
\caption{The geometrical model of the transmit and the receive UCAs in a general non-parallel misalignment case.}
\label{Fig2}
\end{figure}

\subsection{Channel Model}
In free space communications, propagation through the RF channel leads to attenuation and phase rotation of the transmitted signal. This effect is modelled through multiplying by a complex constant $h$, whose value depends on the distance $d$ between the transmit and receive antennas and the wave number $k_p$ \cite{Edfors2012Is}:
\begin{equation} \label{FreeSpaceChannel}
h(d,k_p)=\frac{\beta}{2k_pd}\exp\left(-jk_pd\right),
\end{equation}
where $1/(2k_pd)$ denotes the degradation of amplitude, $\beta$ models all constants relative to the antenna elements and their patterns, and the complex exponential term is the phase difference due to the propagation distance.

According to Fig. \ref{Fig2}, the coordinate of the $n$th antenna element on the transmit UCA is $(R_t \cos\varphi_n, R_t \sin\varphi_n, 0)$ in $Z\!\!-\!\!XOY$ coordinate system, and the coordinate of the $m$th antenna element on the receive UCA in $Z\!\!-\!\!XOY$ coordinate system, denoted by $(B_x,B_y,B_z)$, can be expressed as
\begin{equation}\label{Bxyz}
\left\{
\begin{aligned}
B_x &= R_r\cos\varphi_m\cos\gamma + R_r\sin\varphi_m\sin\alpha\sin\gamma \\
B_y &= R_r\sin\varphi_m\cos\alpha \\
B_z &= D - R_r\cos\varphi_m\sin\gamma + R_r\sin\varphi_m\sin\alpha\cos\gamma
\end{aligned},
\right.
\end{equation}
which is proved in Appendix A. Thus, the transmission distance $d_{m,n}$ from the $n$th ($1\le n \le N$) transmit antenna element to the $m$th ($1\le m \le N$) receive antenna element can be calculated as
\begin{align} \label{dmn}
d_{m,n}&=\left[\left(B_x-R_t\cos\varphi_n\right)^2+\left(B_y-R_t\sin\varphi_n\right)^2+B_z^2\right]^{\frac{1}{2}} \nonumber\\
&=\bigg[R_r^2+R_t^2+D^2-2R_tR_r \sin\varphi_m \cos\varphi_n \sin\alpha \sin\gamma  \nonumber\\
&- 2R_rR_t\left(\cos\varphi_m \cos\varphi_n \cos\gamma + \sin\varphi_m \sin\varphi_n\cos\alpha\right) \nonumber\\
&+ 2DR_r\left(-\cos\varphi_m \sin\gamma + \sin\varphi_m \sin\alpha \cos\gamma \right)
\bigg]^{\frac{1}{2}},
\end{align}
where $D$ is the distance between the transmit and receive UCA centers, $R_t$ and $R_r$ are respectively the radiuses of the transmit and receive UCAs, $\varphi_n=[2\pi(n-1)/N+\varphi_t]$ and $\varphi_m=[2\pi(m-1)/N+\varphi_r]$ are respectively the azimuthal angles of the transmit and receive UCAs, $\varphi_t$ and $\varphi_r$ are the corresponding initial angles of the first reference antenna element in both UCAs, $\alpha$ and $\gamma$ are the oblique angles of the receive UCA as shown in Fig. \ref{Fig2}. For easier analysis, we assume $\varphi_t=0$ and $\varphi_r=0$ here.

Assuming that the transmit and the receive UCAs are placed in the far-field distance of each other, i.e. $D\gg R_t$ and $D\gg R_r$, thus we can approximate $d_{m,n}$ in \eqref{dmn} as
\begin{align} \label{dmn_appx}
d_{m,n}\overset{(a)}{\approx}& \sqrt{R_r^2+R_t^2+D^2}-\frac{R_tR_r \sin\varphi_m \cos\varphi_n \sin\alpha \sin\gamma}{\sqrt{R_r^2+R_t^2+D^2}}\nonumber\\
& -\frac{R_rR_t\left(\cos\varphi_m \cos\varphi_n \cos\gamma + \sin\varphi_m \sin\varphi_n\cos\alpha\right)}{\sqrt{R_r^2+R_t^2+D^2}}\nonumber\\
& +\frac{DR_r\left(-\cos\varphi_m \sin\gamma + \sin\varphi_m \sin\alpha \cos\gamma \right)}{\sqrt{R_r^2+R_t^2+D^2}}\nonumber\\
\overset{(b)}{\approx}& D - \frac{R_rR_t}{D} \sin\varphi_m \cos\varphi_n \sin\alpha \sin\gamma -\frac{R_rR_t}{D}\big( \nonumber\\
& \cos\varphi_m \cos\varphi_n \cos\gamma + \sin\varphi_m \sin\varphi_n\cos\alpha \big)  \nonumber\\
&+R_r \left(-\cos\varphi_m \sin\gamma + \sin\varphi_m \sin\alpha \cos\gamma \right),
\end{align}
where (a) uses the method of completing the square and the condition $D\gg R_t,R_r$ as same as the simple case $\sqrt{a^2-2b}\approx a-\frac{b}{a},a\gg b$, (b) is directly obtained from the condition of $D\gg R_t,R_r$. Then, substituting \eqref{dmn_appx} into \eqref{FreeSpaceChannel} and abbreviating $h(d_{m,n},k_p)$ to $h_{m,n}(k_p)$, we can obtain the channel coefficient from the $n$th transmit antenna element to the $m$th receive antenna element as
\begin{align} \label{hmn}
&h_{m,n}(k_p)\overset{(c)}{\approx}\!\frac{\beta}{2k_p D}\exp\bigg(\!jk_p\frac{R_rR_t}{D}\sin\varphi_m \cos\varphi_n \sin\alpha \sin\gamma  \nonumber\\
&+ jk_p\frac{R_rR_t}{D}\left(\cos\varphi_m \cos\varphi_n \cos\gamma + \sin\varphi_m \sin\varphi_n\cos\alpha \right) \nonumber\\
&- jk_pR_r \left( -\cos\varphi_m \sin\gamma + \sin\varphi_m \sin\alpha \cos\gamma \right) -jk_pD \bigg),
\end{align}
where (c) neglects a few minor terms in the denominator of the amplitude term and thus only $2k_pD$ is left.
In the end, the channel matrix of the UCA-based free space OAM communication system can be expressed as $\mathbf{H}(k_p)=[h_{m,n}(k_p)]_{N\times N}$.

\section{The Effect of Oblique Angles on the Performance of Multi-mode OAM Broadband Communication Systems}
In this section, we analyze the effect of oblique angles on the performance of an ideal multi-mode OAM broadband system from the perspective of channel gain and IMI.

\subsection{The Effect of $\alpha$ and $\gamma$ on channel gain and IMI }\label{effect}
In \cite{Zhang2017Mode}, the OAM channel is defined as $\mathbf{H}_{\mathrm{OAM},k_p}=\mathbf{F}_U \mathbf{H}(k_p)\mathbf{F}_U^H$, which will become a diagonal matrix $\mathbf{\Lambda}_U(k_p)$ due to the circularity of $\mathbf{H}(k_p)$ when $\alpha=0$ and $\gamma=0$. However, in the case of misalignment, $\mathbf{H}(k_p)$ is not a circulant matrix any more due to the effect of $\alpha$ and $\gamma$.

In the misalignment case, the $u$th-row and $v$th-column element of $\mathbf{H}_{\mathrm{OAM},k_p}$ can be calculated as
\begin{align} \label{h_OAM}
&h_{\mathrm{OAM},k_p}(u,v)=\mathbf{f}(\ell_u) {\mathbf{H}(k_p)} \mathbf{f}^H(\ell_v) \nonumber\\
&=\frac{1}{N}\sum\limits_{m = 1}^N \sum\limits_{n = 1}^N h_{m,n}(k_p) \exp \left(-j\ell_u \varphi_m + j\ell_v \varphi_n \right) \nonumber\\
&\overset{(a)}\approx \eta(k_p) \sum\limits_{m = 1}^N \sum\limits_{n = 1}^N \exp \bigg( jk_p\frac{R_rR_t}{D} \cos(\varphi_n-\varphi_m) \nonumber\\
&\quad -j k_pR_r(-\cos\varphi_m \sin\gamma + \sin\varphi_m \sin\alpha \cos\gamma)  \nonumber\\
&\quad + j\ell_v\left(\varphi_n-\varphi_m\right) -j\varphi_m(\ell_u-\ell_v) \bigg) \nonumber\\
&=\eta(k_p)\sum\limits_{s = 1}^N \exp \bigg(jk_p\frac{R_rR_t}{D}\cos\frac{2\pi s}{N} + j\frac{2\pi s}{N}\ell_v \bigg) \nonumber\\
&\quad \times \xi(\alpha,\gamma,t) ,
\end{align}
where $\eta(k_p)=\frac{\beta}{2k_pDN}\exp(-jk_pD)$, $s=n-m\in \mathbb{Z}$, $t=\ell_{u}-\ell_{v}\in \mathbb{Z}$ refers to the difference between the two OAM modes, (a) applies
the approximation $\cos\alpha\approx1-\frac{\alpha^2}{2}$ for $\cos\alpha$ and $\cos\gamma\approx1-\frac{\gamma^2}{2}$ for $\cos\gamma$ in the case that $\alpha$ and $\gamma$ are relatively small and neglects a few quadratic terms, and
\begin{align}\label{xi}
\xi(\alpha,\gamma,t)=&\sum\limits_{m = 1}^N \exp \big(-j k_p R_r(-\cos\varphi_m \sin\gamma  \nonumber\\
&+ \sin\varphi_m \sin\alpha \cos\gamma) -j\varphi_mt \big).
\end{align}
For easier expression, we refer to $|{h_{\mathrm{OAM},k_p}(u,u)}|^2$ as the channel gain, and $\sum\limits_{v \ne u} |{ h_{\mathrm{OAM},k_p}(u,v)}|^2$ as the IMIs. From \eqref{h_OAM} and \eqref{xi}, we can observe that, when $\alpha\neq0$ or $\gamma\neq0$ and $t\neq0$, then $\xi\neq0$, which indicates that $\alpha$ and $\gamma$ result in IMIs and thus ${{\bf{H}}_{\textrm{OAM},k_p}}$ can not be transformed into a diagonal matrix in the misalignment case.

For further study, we let $\sin \gamma =\nu$, $\sin \alpha \cos\gamma = \mu$, then \eqref{xi} can be written as
\begin{flalign}\label{xi1}
\xi(\alpha,\gamma,t)\!=&\! \sum\limits_{m = 1}^N \! \! \exp \! \left(-j k_p R_r(-\nu \cos\varphi_m \! + \! \mu \sin\varphi_m )\! -\!j\varphi_mt \right) \nonumber \\
\!=&\!\sum\limits_{m = 1}^N \! \! \exp \left(-j k_p R_r\rho \sin(\varphi_m - \phi) -j\varphi_mt \right),
\end{flalign}
where $\rho=\sqrt{\mu^2+\nu^2}$, $\phi=\arctan\frac{\nu}{\mu}$. Based on the similar treatment in \cite{Mohammadi2010Orbital}, when $N$ is large enough, \eqref{h_OAM} can be approximated by an integral as
\begin{align} \label{h_OAM1}
&h_{\mathrm{OAM},k_p}(u,v) \nonumber\\
&\approx  \eta(k_p) \frac{N}{2\pi} \!  \int_0^{2\pi} \! \exp \bigg(jk_p\frac{R_rR_t}{D}\cos x + jx\ell_v \bigg)dx \nonumber\\
&\quad \times \frac{N}{2\pi} \! \! \int_0^{2\pi} \! \! \exp \bigg(\! \! -jk_pR_r\rho\sin(x-\phi)\! - \! jxt \bigg)dx \nonumber\\
&\approx  \eta'(k_p) J_{\ell_v} \left(k_p\frac{R_rR_t}{D}\right) J_{t}\big(k_pR_r\rho),
\end{align}
where $\eta'(k_p)=\eta(k_p)N^2\exp(j\frac{\pi}{2}\ell_v-j\pi t - j\phi t)=\frac{N\beta}{2k_pD}\exp($
$-jk_pD+j\frac{\pi}{2}\ell_v-j\pi t-j\phi t)$, $J_{n}(\cdot)$ represents $n$-order Bessel function of the first kind.
According to the characteristic of Bessel function, it's easy to obtain that, when $t\neq0$, if $\rho\neq0$ ($\alpha\neq0$ or $\gamma\neq0$), $J_{t}\big(k_pR_r\rho)\neq0$, which will result in large IMIs even if $\alpha$ and $\gamma$ have small values due to large $k_p$. Moreover, when $t=0$, we can find that if $\rho=0$, $J_{t}\big(k_pR_r\rho)=1$, if $\rho\neq0$, $J_{t}\big(k_pR_r\rho)$ will be much less than 1, which indicates that $\alpha$ and $\gamma$ will result in much loss of channel gain.

\subsection{Analysis of SINR}
The received modulation symbols at the $u$th mode and $p$th subcarrier at the receiver in the misalignment case, denoted as $y(k_p,\ell_u)$, can be written as
\begin{align} \label{yM}
&y(k_p,\ell_u)= \sum\limits_{v = 1}^U {{h_{\mathrm{OAM},k_p}(u,v)}{x(k_p,\ell_v)} + {\tilde{z}(k_p,\ell_u)}}  \nonumber\\
&= {h_{\mathrm{OAM},k_p}(u,u)}{x(k_p,\ell_u)} + \sum\limits_{v \ne u} {h_{\mathrm{OAM},k_p}(u,v)}{x(k_p,\ell_v)}  \nonumber\\
&+ \tilde{z}(k_p,\ell_u),
\end{align}
where ${h_{\mathrm{OAM},k_p}(u,u)}{x(\ell_u)}$ is the useful signal information, $\sum\limits_{v \ne u}^{} {{h_{\mathrm{OAM},k_p}(u,v)}{x(\ell_v)}}$ is the interference information, $\tilde{z}(k_p,\ell_u)=\mathbf{f}({\ell_u})\mathbf{z}(k_p)$ is also a complex Gaussian variable with zero mean and variance $\sigma_z^2$ since $\mathbf{f}(\ell_u)$ doesn't change the noise power.

Therefore, the signal-to-interference-plus-noise ratio (SINR) at the $u$th mode and $p$th subcarrier can be formulated as
\begin{align} \label{SINR}
\textrm{SINR}_{u}(k_p)&= \frac{{{{\left| {{h_{\mathrm{OAM},k_p}(u,u)}} \right|}^2}\mathbb{E}\left( {{{\left| {{x(k_p,\ell_u)}} \right|}^2}} \right)}}{{\sum\limits_{v \ne u} {{{\left| {{h_{\mathrm{OAM},k_p}(u,v)}} \right|}^2}\mathbb{E}\left( {{{\left| {{x(k_p,\ell_v)}} \right|}^2}} \right)}  + \sigma^2_z}},
\end{align}
where $\mathbb{E}\left( {{{\left| {{x(k_p,\ell_u)}} \right|}^2}} \right)$ is the average power of modulation symbol at the $u$th mode and $p$th subcarrier.

Additionally, from \eqref{h_OAM1} and the expression of $\eta'(k_p)$, we can observe that, when other parameters are fixed, the value of $|h_{\mathrm{OAM},k_p}(u,v)|$ linearly increases with $N$, which means both channel gain and IMIs linearly increase with $N$. In order to analyze the effect of large $N$ on the SINR, we define $\tilde{h}_{\mathrm{OAM},k_p}(u,v) = h_{\mathrm{OAM},k_p}(u,v)/N$. Then, $\textrm{SINR}_u(k_p)$ can be calculated as
\begin{align} \label{SINR_WoBS2}
\textrm{SINR}_u(k_p) &= \frac{{\left| \tilde{h}_{\mathrm{OAM},k_p}(u,u) \right|}^2\mathbb{E}\left( {{{\left| {{x(k_p,\ell_u)}} \right|}^2}} \right)}{{\sum\limits_{v \ne u} {{{\left| \tilde{h}_{\mathrm{OAM},k_p}(u,v) \right|}^2}\mathbb{E}\left( {{{\left| {x(k_p,\ell_v)} \right|}^2}} \right)}  + \frac{\sigma_z^2}{N^2}}} \nonumber \\
&\overset{(a)}\approx \textrm{SIR}_u(k_p),
\end{align}
where (a) holds when $N$ is large enough, and $\textrm{SIR}_u(k_p)$ is defined as the signal-to-interference ratio (SIR) at the $u$th mode and $p$th subcarrier. Hence, using large number of antennas can eliminate the uncorrelated noise, but has no effect on the IMIs of the misaligned OAM communication system, which is different from the massive MIMO technology being able to alleviate the interference between multiple data streams \cite{Marzetta2010Noncooperative}. It follows that interference cancellation technology, such as beam steering, is necessary for a practical OAM communication system.

Moreover, it's obvious that $\textrm{SINR}_{u}(k_p)$ is the function of subcarrier frequency as shown in \eqref{SINR}. In order to explain it intuitively, we plot the relationship between SINR at the $u$th mode and subcarrier frequency in Fig. \ref{Fig3} with the system parameters specified in Table \ref{Tab1}, where $\ell_u=1$, and $\alpha=\gamma=1^{\circ}, 5^{\circ}, 10^{\circ}, 15^{\circ}, 20^{\circ}$. In Fig. \ref{Fig3}, we can observe that the SINR values decrease with the increase of the oblique angle, and for a fixed oblique angle the SINRs vary with the subcarrier frequency, which indicates that the phase compensation through phase substraction \cite{Chen2017Misalignment} or beam steering \cite{Chen2018Beam} for the received signals at different subcarriers should be distinct.

\begin{figure}[t]
\begin{center}
\includegraphics[width=8cm]{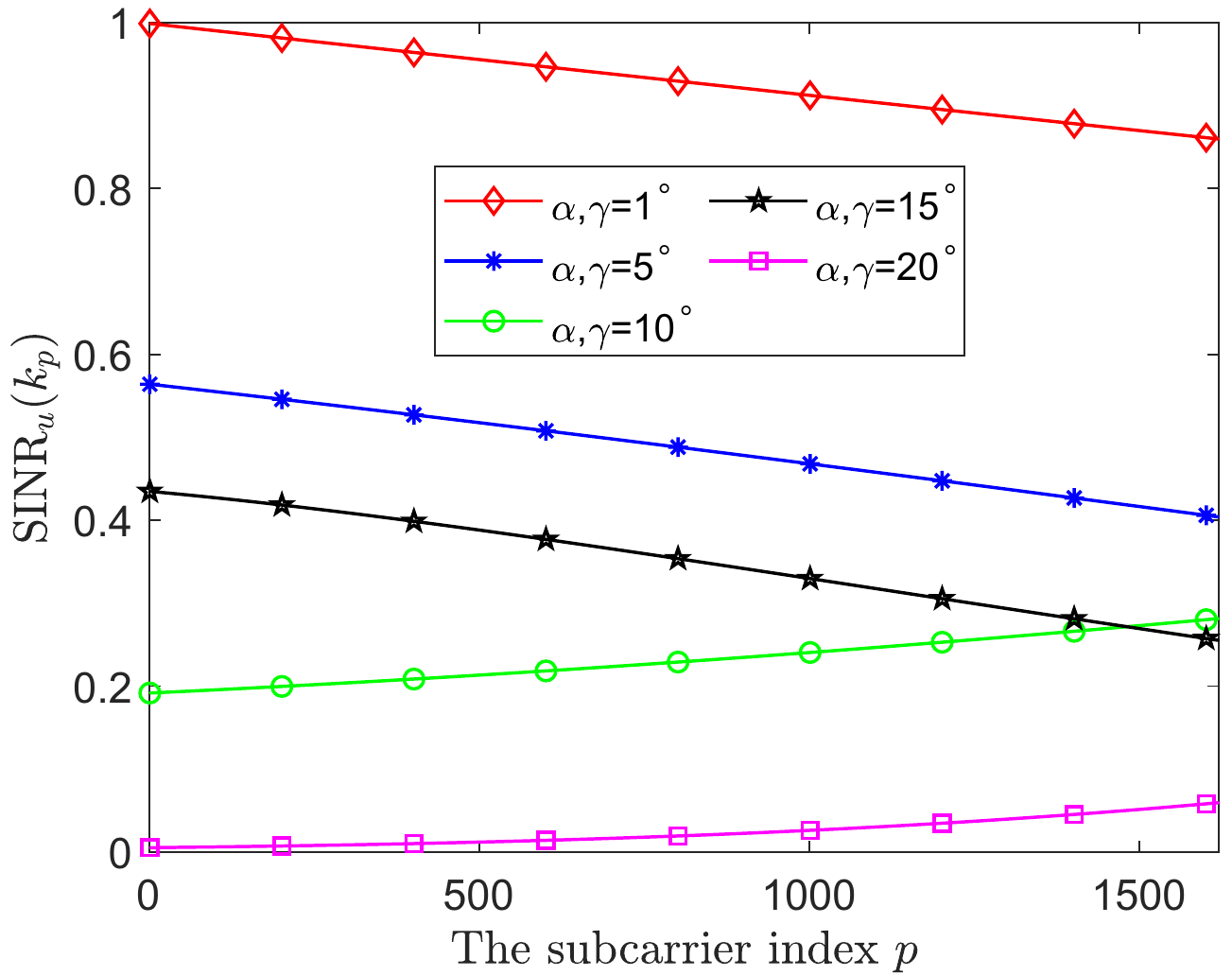}%
\end{center}
\caption{The relationship between SINR and subcarrier frequency.}
\label{Fig3}
\end{figure}
\renewcommand{\arraystretch}{1.5}
\begin{table}[tb]
  \centering 
  \caption{System Parameters for the multi-mode OAM broadband system}\label{Tab1}
\begin{tabular}{ccc}
  \hline
  \hline
  Parameters & values  \\
  \hline

  Antenna number of UCA [$N$]  & 15 \\

  The number of OAM modes [$U$]  & 15 \\

  Content for channel [$\beta$ (dB)] & 24.7 \\

  Center frequency [$f_c=c/\lambda_0$ (GHz)] & 3.5 \\

  The radius of transmit UCA [$R_t$]  & 10$\lambda_0$ \\

  The radius of receive UCA [$R_r$]  & 10$\lambda_0$ \\

  The distance between two UCAs [$D$] & 300$\lambda_0$ \\

  Bandwidth [$B$] (MHz) & 100  \\

  The number of subcarriers [$P$] & 1620  \\

  Subcarrier spacing [$\Delta f$ (kHz)] & 60 \\

  The lowest frequency of system [$f_L$] & $f_c-\frac{P}{2}\times\Delta f$\\

  The $p$th subcarrier for channel [$f_p$] & $f_L+p\times\Delta f$\\

%

  SNR (dB) & 20 \\

  Noise power [$\sigma^2_z$] & 0.01 \\

  \hline
  \hline
\end{tabular}
\end{table}
%


\section{Beam Steering for the Misaligned Multi-Mode OAM Broadband System}
\label{Beam_Steering}
To alleviate the effect of oblique angles on the channel gain and IMIs induced by the misalignment of transmit and receive UCAs, we consider applying beam steering to the misaligned multi-mode OAM broadband communication system. Beam steering can steer the beam pattern towards the direction of the incident OAM beam and thus compensate the changed phases caused by $\alpha$ and $\gamma$ at the receive UCA. We consider two beam steering schemes, i.e. ABS and DBS, and compare their differences in this section.

\subsection{Analog Beam Steering}
Through calculating the phase difference between the reference element and the $m$th element of the receive UCA, the beam steering matrix of ABS at the receiver $\mathbf{W}_A$ can be written as
\begin{equation} \label{Wa}
\mathbf{W}_A =\mathbf{1}_U\otimes\mathbf{w}_A,
\end{equation}
where $\mathbf{w}_A=[ e^{jw^A_1}, e^{jw^A_2}, \cdots, e^{jw^A_N}]$,
%
\begin{equation} \label{w_m}
w^A_m=k_AR_r\left(-\cos\varphi_m\sin\gamma + \sin\varphi_m\sin\alpha\cos\gamma\right),
\end{equation}
$m=1,\cdots,N$, $k_A=\frac{2\pi}{c}f_A$, $f_A=f_L+A\times \Delta f$ is the fixed frequency of RF analog receiver, $f_L$ is the lowest frequency of the broadband system, $\Delta f$ is the subcarrier spacing, $A\in\{1,2,\cdots,P\}$.
After involving these phases into the original phases in ${\bf{F}}_U$ at the receive UCA, the effective multi-mode OAM channel after ABS at the receiver can be expressed as
\begin{equation} \label{H_OAM_ABS}
{{\mathbf{H}}^{\mathrm{ABS}}_{\textrm{OAM},k_p}} = \left( {\mathbf{W}_A} \odot {{\mathbf{F}}_U} \right){\mathbf{H}(k_p)\mathbf{F}}_U^H.
\end{equation}
\begin{thm} \label{Thm1}
For a multi-mode OAM broadband communication system composed of an N-elements transmit UCA and an N-elements receive UCA, after ABS, there still remain IMIs induced by $\alpha$ and $\gamma$ in a non-parallel misalignment case.
\end{thm}
\begin{IEEEproof}
The $u$th-row and $v$th-column element in matrix ${{\mathbf{H}}^{\mathrm{ABS}}_{\textrm{OAM},k_p}}$ can be obtained as
\begin{align} \label{h_OAM_ABS}
&h^{\mathrm{ABS}}_{\textrm{OAM},k_p}(u,v) =  \left(\mathbf{w}_A \odot \mathbf{f}(\ell_u)\right) \mathbf{H}(k_p) \mathbf{f}^H( {\ell_v} ) \nonumber\\
&\approx \frac{1}{N} \sum\limits_{n = 1}^N \sum\limits_{m = 1}^N h_{m,n}(k_p) \exp\bigg(-j\ell_u \varphi_m + j\ell_v \varphi_n \nonumber\\
&\quad +jk_AR_r\left(-\cos\varphi_m\sin\gamma + \sin\varphi_m\sin\alpha\cos\gamma\right) \bigg) \nonumber\\
&= \eta(k_p)\sum\limits_{s = 1}^N \exp \bigg(jk_p\frac{R_rR_t}{D}\cos\frac{2\pi s}{N} + \frac{2\pi s}{N}\ell_v \bigg) \nonumber\\
&\quad\times \sum\limits_{m = 1}^N \exp \left(-j k_g R_r(\rho \sin(\varphi_m-\phi)-j\varphi_mt \right)\nonumber\\
&\approx \eta'(k_p) J_{\ell_v} \left(k_p\frac{R_rR_t}{D}\right)J_{t}\big(k_{g}R_r\rho),
\end{align}
where $k_g=k_p-k_A=\frac{2\pi}{c}(f_p-f_A)=\frac{2\pi}{c}g\Delta f$, $g=p-A$, and $g\Delta f$ represents the frequency deviation between channel at the $p$th subcarrier and RF analog receiver. In order to guarantee the minimum whole frequency deviation of the OAM broadband system, we choose $A=\frac{P}{2}$, namely, $f_A=f_c$, where $f_c$ is the center frequency of system. From \eqref{h_OAM_ABS}, we can see that there still remain IMIs after ABS and it's value depends on $k_gR_r\rho$.
\end{IEEEproof}

Based on the above analysis, after ABS, the SINR at the $u$th mode and $p$th subcarrier can be formulated as
\begin{align} \label{SINR_ABS}
&\textrm{SINR}^{\mathrm{ABS}}_{u}(k_p) = \frac{{{{\left| {{h^{\mathrm{ABS}}_{\mathrm{OAM},k_p}(u,u)}} \right|}^2}\mathbb{E}\left( {{{\left| {{x(k_p,\ell_u)}} \right|}^2}} \right)}}{{\sum\limits_{v \ne u} {{{\left| {{h^{\mathrm{ABS}}_{\mathrm{OAM},k_p}(u,v)}} \right|}^2}\mathbb{E}\left( {{{\left| {{x(k_p,\ell_v)}} \right|}^2}} \right)}  + \sigma^2_z}}.
\end{align}
\begin{rem} \label{Remark1}
It is revealed in \eqref{h_OAM_ABS} that the value of $|J_{t}\big(k_{g}R_r\rho)|$ will increase with $k_g$. Thus, once the bandwidth of OAM system is substantially increased for future applications, the IMIs after ABS will become larger because of the increased frequency deviation. That is to say, the IMIs induced by ABS become more serious in the OAM system with larger bandwidth.
\end{rem}
\subsection{Digital Beam Steering}
As for DBS, the receiver beam steering matrix at the $p$th subcarrier $\mathbf{W}_D(k_p)$ can be calculated as
\begin{equation} \label{W}
\mathbf{W}_D(k_p) =\mathbf{1}_U\otimes\mathbf{w}_D(k_p),
\end{equation}
where $\mathbf{w}_D(k_p)=[ e^{jw^D_1(k_p)}, e^{jw^D_2(k_p)}, \cdots, e^{jw^D_N}(k_p)]$,
\begin{equation} \label{w_m}
w^D_m(k_p)=k_pR_r\left(-\cos\varphi_m\sin\gamma + \sin\varphi_m\sin\alpha\cos\gamma\right),
\end{equation}
and $m=1,\cdots,N$. After involving these phases into the original phases in ${\bf{F}}_U$ at the receive UCA, the effective OAM channel matrix after DBS at the receiver becomes
\begin{equation} \label{eq17}
{{\mathbf{H}}^{\mathrm{DBS}}_{\textrm{OAM},k_p}} = \left( {\mathbf{W}_D(k_p)} \odot {{\mathbf{F}}_U} \right){\mathbf{H}(k_p)\mathbf{F}}_U^H.
\end{equation}
\begin{thm} \label{Thm2}
For a multi-mode OAM broadband communication system composed of an N-elements transmit UCA and an N-elements receive UCA, beam steering matrix of DBS $\mathbf{W}_D(k_p)$ can eliminate IMIs induced by $\alpha$ and $\gamma$ in a non-parallel misalignment case.
\end{thm}
\begin{IEEEproof}
The $u$th-row and $v$th-column element in matrix ${{\mathbf{H}}^{\mathrm{DBS}}_{\textrm{OAM},k_p}}$ can be obtained as
\begin{align} \label{heff_DBS}
&h^{\mathrm{DBS}}_{\textrm{OAM},k_p}(u,v) = \left(\mathbf{w}_D(k_p) \odot \mathbf{f}(\ell_u) \right) \mathbf{H}(k_p) \mathbf{f}^H( {\ell_v} ) \nonumber\\
&\approx \frac{1}{N} \sum\limits_{n = 1}^N \sum\limits_{m = 1}^N h_{m,n}(k_p) \exp\bigg(-j\ell_u \varphi_m + j\ell_v \varphi_n \nonumber\\
&\quad +jk_pR_r\left(-\cos\varphi_m\sin\gamma + \sin\varphi_m\sin\alpha\cos\gamma\right) \bigg) \nonumber\\
& = \eta(k_p)\sum\limits_{s = 1}^N \exp \bigg(jk_p\frac{R_rR_t}{D}\cos\frac{2\pi s}{N} + \frac{2\pi s}{N}\ell_v \bigg) \nonumber\\
&\quad\times \sum\limits_{m = 1}^N \exp \left(-j\varphi_mt \right)  \nonumber\\
&\approx
\begin{cases}
\eta''(k_p) J_{\ell_v} \big(k_p\frac{R_rR_t}{D}), & t=0 \ (u=v)\\
0 \qquad \qquad \qquad \quad \ \ , & t\neq 0 \ (u\neq v),
\end{cases}
\end{align}
where $\eta''(k_p)=\frac{N\beta}{2k_pD}\exp(-jk_pD+j\frac{\pi}{2}\ell_v)$.
It's obvious that the value of $h^{\mathrm{DBS}}_{\textrm{OAM},k_p}(u,v)$ is not sensitive to oblique angles. Specifically, when $t\neq0$, even through $\alpha\neq0$ and $\gamma\neq0$, $h^{\mathrm{DBS}}_{\textrm{OAM},k_p}(u,v)\approx0$. That is to say, the non-diagonal elements of the effective multi-mode OAM channel matrix after DBS becomes vanishing small, which proves that DBS can eliminate the IMIs almost completely.
\end{IEEEproof}

Thus, after DBS, the SINR at the $u$th mode and $p$th subcarrier can be formulated as
\begin{align} \label{SINR_DBS}
&\textrm{SINR}^{\mathrm{DBS}}_{u}(k_p)\approx \frac{{{{\left| {{h^{\mathrm{DBS}}_{\mathrm{OAM},k_p}(u,u)}} \right|}^2}\mathbb{E}\left( {{{\left| {{x(k_p,\ell_u)}} \right|}^2}} \right)}}{\sigma^2_z}.
\end{align}

\subsection{Applicability Analysis of ABS and DBS}
Comparing \eqref{h_OAM1} and \eqref{h_OAM_ABS}, we can find that the difference between the misaligned OAM channel element and the effective channel element after ABS is the second Bessel function term, and their values are respectively dependent on the size of $k_{p}R_r\rho$ and $k_{g}R_r\rho$. In order to explain it intuitively, we give the specific values of $k_{p}R_r\rho$ and $k_{g}R_r\rho$ in Table \ref{Tab2} with the system parameters specified in Table \ref{Tab1}.
It's clear that $k_{g}R_r\rho \ll k_{p}R_r\rho$ as shown in Table \ref{Tab2}. According to the characteristics of Bessel function, we can infer that when $t=0$, $|J_{t}\big(k_{g}R_r\rho)|$ will be much larger than $|J_{t}\big(k_{p}R_r\rho)|$, and when $t\neq0$, $|J_{t}\big(k_{g}R_r\rho)|$ will be much lower than $|J_{t}\big(k_{p}R_r\rho)|$ if oblique angles are relatively small. Therefore, compared with misalignment case, ABS can obtain much larger channel gain and much lower IMIs, and thus effectively alleviate the destructive effect of oblique angles on the transmission performance of the misaligned multi-mode OAM broadband system. However, it's obvious that $k_{g}R_r\rho$ increases with oblique angles and $|J_{t}\big(k_{g}R_r\rho)|$ becomes large when oblique angles are relatively large (e.g., $\alpha,\gamma>10^{\circ}$), which indicates that there remains large IMIs even after ABS.
\newcommand{\tabincell}[2]{\begin{tabular}{@{}#1@{}}#2\end{tabular}}  
\begin{table}[tb]
  \centering 
  \caption{The values of $k_{p}R_r\rho$ and $k_{g}R_r\rho$}\label{Tab2}
\begin{tabular}{cccc}
  \hline
  \hline
      & $k_{p}R_r\rho$  & $k_{g}R_r\rho$  \\
  \hline
  $\alpha,\gamma=1^{\circ}$ & $1.55$  & $2.15\times10^{-2}$ \\

  $\alpha,\gamma=5^{\circ}$ & $7.73$ & $1.07\times10^{-1}$ \\

  $\alpha,\gamma=10^{\circ}$ & $15.31$  & $2.13\times10^{-1}$\\

  $\alpha,\gamma=15^{\circ}$ & $22.6$  & $3.14\times10^{-1}$ \\

  $\alpha,\gamma=20^{\circ}$ & $29.49$  & $4.09\times10^{-1}$  \\
  \hline
  \hline
\end{tabular}
\end{table}

In order to explain it more directly, we give a toy example in Table \ref{Tab3} with the system parameters specified in Table \ref{Tab1}, where MA represents the misalignment case, $\ell_u=1$, $\alpha=\gamma= 5^{\circ}, 15^{\circ}$, and $f_p=f_L$. We can observe that the results in Table \ref{Tab3} are coincident with above analyses. Especially, when $\alpha=\gamma=15^{\circ}$, the IMI of DBS is so small that can be neglected, while the IMI of ABS is comparable to the noise power, and thus decreases the system performance to a certain extent compared with DBS. Therefore, although ABS can improve the system performance, DBS is the better choice for the misaligned multi-mode OAM broadband system.
\begin{table}[tb]
\centering 
\caption{Comparison of channel gain and IMI between misalignment case, ABS and DBS.}
\label{Tab3}
\begin{tabular}{ccc}
\hline
\hline
&Channel gain & IMI \\
\hline

MA ($\alpha,\gamma=5^{\circ}$) & $2.99\times10^{-2}$   & $4.29\times 10^{-2}$ \\ 

ABS ($\alpha,\gamma=5^{\circ}$) & $4.76\times10^{-1}$   & $8.52\times 10^{-4}$ \\ 

DBS ($\alpha,\gamma=5^{\circ}$) & $4.79\times10^{-1}$ & $2.57\times 10^{-5}$  \\ 

MA ($\alpha,\gamma=15^{\circ}$) & $2.74\times10^{-2}$   & $5.30\times 10^{-2}$ \\ 

ABS ($\alpha,\gamma=15^{\circ}$) & $4.58\times10^{-1}$   & $8.11\times 10^{-3}$ \\ 

DBS ($\alpha,\gamma=15^{\circ}$) & $4.85\times10^{-1}$ & $9.72\times 10^{-4}$  \\ 

\hline
\hline
\end{tabular}
\end{table}

\section{Numerical Simulations and Results}
In this section, the practicability of DBS for the misaligned multi-mode OAM broadband system is verified through being compared with perfect alignment, misalignment case and ABS. Except for special explanation, the simulation parameters are listed in Table \ref{Tab1}, and the SNR is defined as the ratio of the transmit power versus the noise power.

First, we show the effect of beam steering on the channel gain and IMIs of the misaligned multi-mode OAM broadband system, where $\alpha$ ranges from $0^{\circ}$ to $20^{\circ}$, $\gamma= 0^{\circ}, 10^{\circ}, 20^{\circ}$, and $f_p=f_L$. For a multi-mode OAM system, we refer to $\mathrm{\overline{CG}}=\frac{1}{U}\sum^U_{u=1}|h_{\mathrm{OAM},k_p}(u,u)|^2$ as the average channel gain, and
$\mathrm{\overline{IMI}}=\frac{1}{U}\sum^U_{u=1}\sum\limits_{v \ne u}|h_{\mathrm{OAM},k_p}(u,v)|^2$ as the average IMI. In Fig. \ref{Fig4}, we can find that $\mathrm{\overline{CG}}$ in the misalignment case decreases rapidly and $\mathrm{\overline{IMI}}$ increases quickly, even though there is a small increment on $\alpha$ and $\gamma$. However, no matter how large $\alpha$ and $\gamma$ are, $\mathrm{\overline{CG}}$s of ABS and DBS still remain in a high value, while $\mathrm{\overline{IMI}}$s are much lower than that of misalignment case. It's obvious that ABS and DBS both can greatly improve the transmission performance of the misaligned multi-mode OAM broadband system.
\begin{figure}[t]
\centering
\subfigure[]
{\includegraphics[width=8cm]{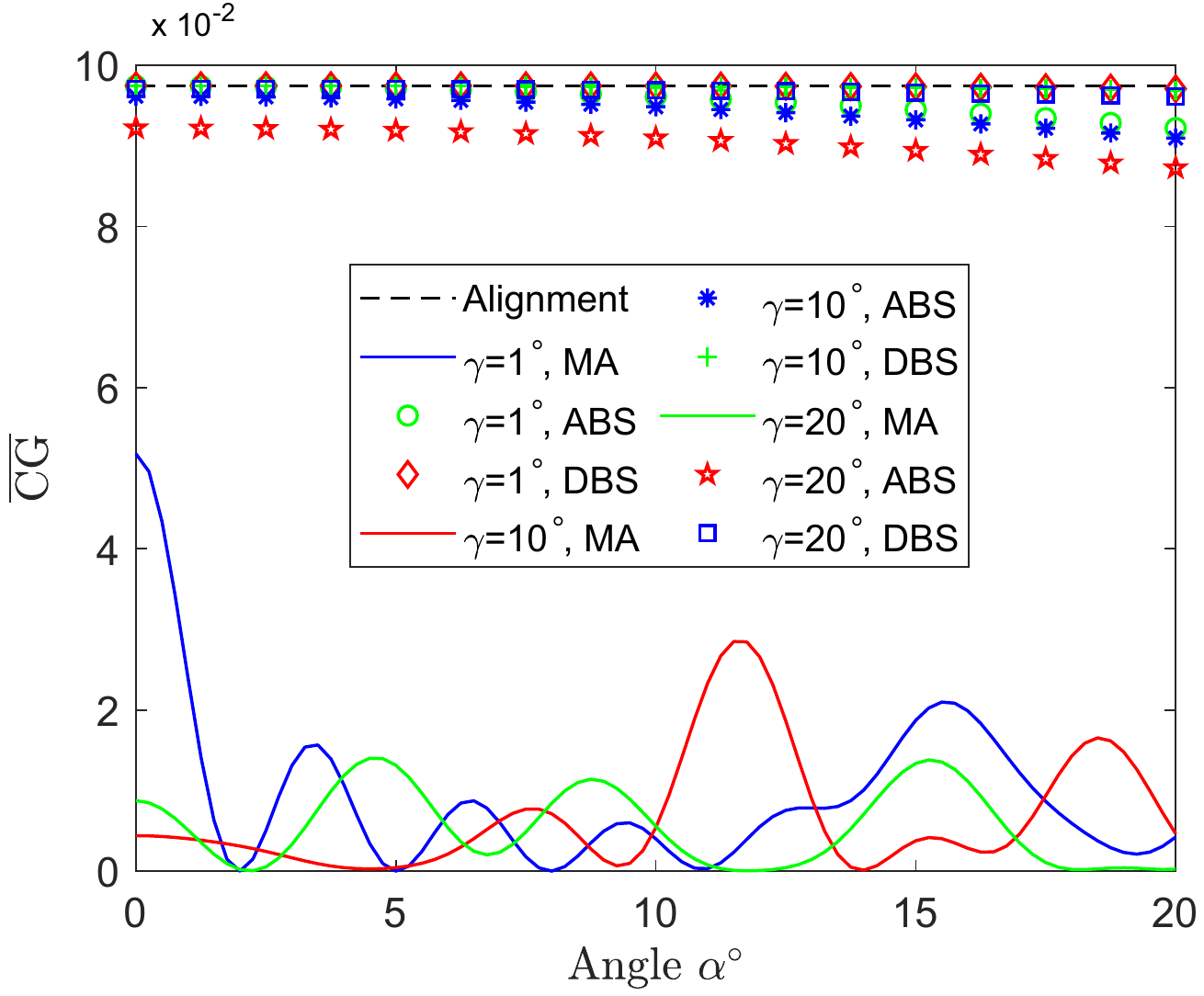}}
\subfigure[]
{\includegraphics[width=8cm]{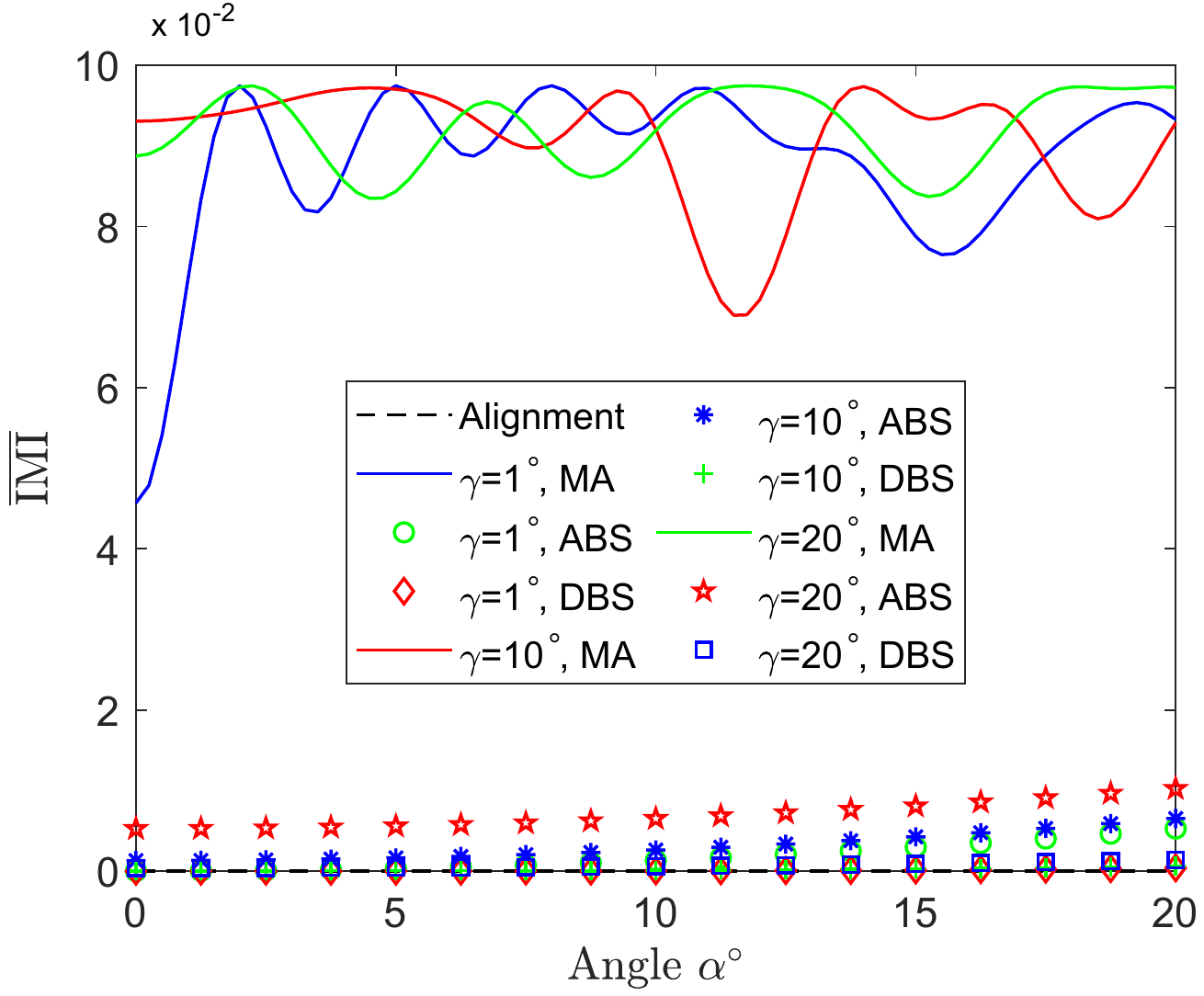}}
\caption{In the misaligned multi-mode OAM broadband system: (a) the average channel gain vs. oblique angles; (b) the average IMI vs. oblique angles.  }
\label{Fig4}
\end{figure}

Then, we compare the performance of ABS and DBS in the misaligned multi-mode OAM broadband system. As shown in Fig. \ref{Fig4}, $\mathrm{\overline{CG}}$ of DBS can remain basically unchanged and is approximately consistent with that of perfect alignment, while $\mathrm{\overline{CG}}$ of ABS has a visible decrease, especially in large oblique angles. Correspondingly, $\mathrm{\overline{IMI}}$ of ABS increases as oblique angles increase but $\mathrm{\overline{IMI}}$ of DBS always remains in a quite low level. Therefore, the transmission performance of the misaligned multi-mode OAM broadband system with ABS will decrease in a certain degree compared with that of DBS.

Thereafter, we calculate SE to measure the overall transmission performance of the misaligned multi-mode OAM broadband system, where $\alpha=\gamma=10^{\circ}, 15^{\circ}, 20^{\circ}$. In Fig.\ref{Fig5}, we can observe that the SE in the misalignment case is so small that the misaligned OAM system without beam steering can't work normally. Both ABS and DBS can greatly improve the SE of the misaligned multi-mode OAM broadband system. However, the SE of ABS continues to decrease as oblique angles increase, and is always lower than that of DBS. Specifically, the maximum SE of ABS is about 3 bit/s/Hz lower than that of DBS, and about 5.5 bit/s/Hz lower than that of perfect alignment when $\alpha, \gamma=20^{\circ}$. This decline cannot be ignored under the serious situation of scarce communication resources. Therefore, benefiting from higher SE, DBS is more practical for the misaligned multi-mode OAM broadband system.

Finally, we analyze the effect of $N$ on the SE of the misaligned multi-mode OAM broadband system, where $\alpha=\gamma=20^{\circ}$, and $N=15, 32$. Fig. \ref{Fig6} illustrates that the SE of misalignment case can't get noticeable improvement even if we increase the number of antenna. Fortunately, after applying ABS and DBS, the SEs get an apparent increment. Moreover, we can find that the difference of SE between ABS and DBS increases with the number of antenna $N$, for example, the difference of SE between ABS and DBS reaches up 6 bit/s/Hz when $N=32$, which verifies DBS is more suitable for the misaligned multi-mode OAM broadband system again, especially in the system with large antennas.


%
\begin{figure}[t]
\begin{center}
\includegraphics[width=8cm]{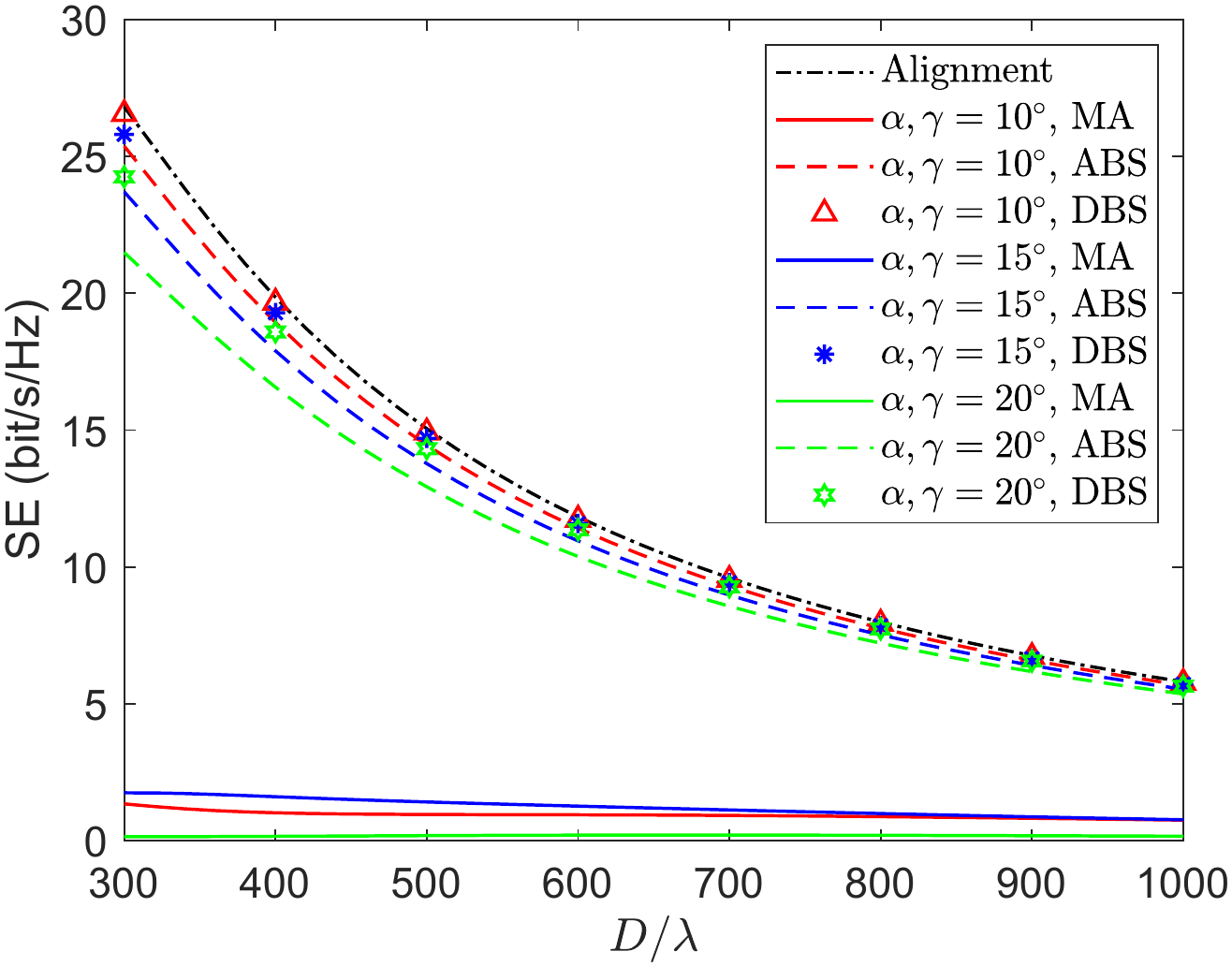}%
\end{center}
\caption{Comparison of SE between perfect alignment, misalignment, ABS and DBS in the misaligned multi-mode OAM broadband system.}
\label{Fig5}
\end{figure}
\begin{figure}[t]
\begin{center}
\includegraphics[width=8cm]{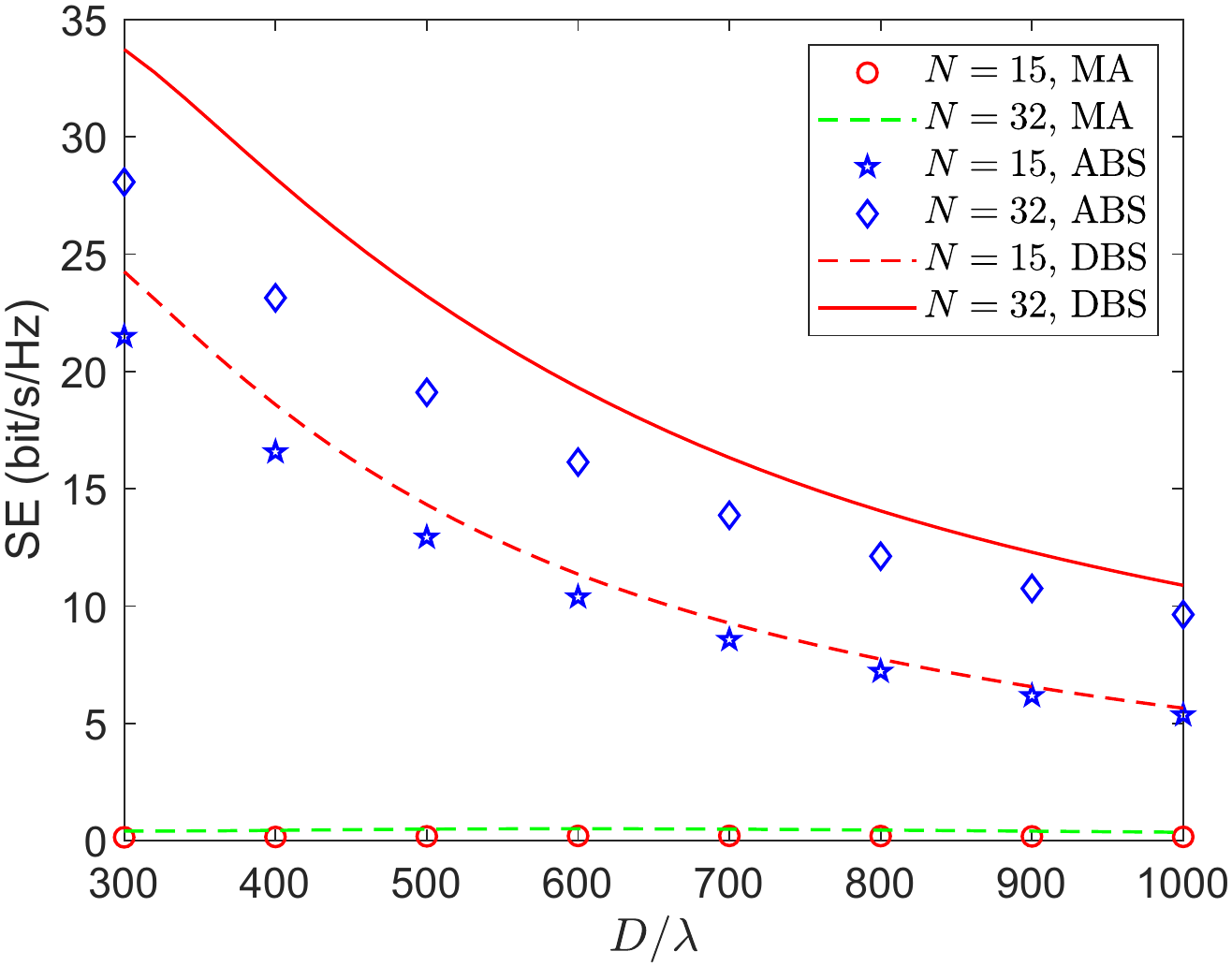}%
\end{center}
\caption{The effect of $N$ on the SE of the misaligned multi-mode OAM broadband system.}
\label{Fig6}
\end{figure}

\section{Conclusions}
In this paper, aiming at the application challenge of UCA-based OAM broadband communication, we first quantitatively investigate the effect of oblique angles on the transmission performance of the multi-mode OAM broadband communication system in a non-parallel misalignment case, which shows the performance of the OAM broadband system decreasing significantly even under small oblique angles, and having little improvement as $N$ increases. Therefore, corresponding to the  RF analog and baseband digital transceiver structures, we adopt two beam steering schemes, i.e., ABS and DBS, to mitigate the destructive effect of the misalignment. Mathematical analysis and numerical simulations validate that ABS and DBS both can greatly improve the SE of the misaligned multi-mode OAM broadband system, but DBS can obtain higher SE than ABS, and the larger the number of antennas, the better the performance. This fact indicates that baseband digital transceiver with DBS is a better option for the multi-mode OAM broadband communication system in practice.

\section*{Appendix A. Derivation of \eqref{Bxyz}}
\renewcommand{\thesection}{\Alph{section}}
\renewcommand{\theequation}{A.\arabic{equation}}
\setcounter{equation}{0}
\setcounter{thm}{1}
As shown in Fig. \ref{Fig2}, the coordinate of the $m$th antenna element on the transmit UCA can be written as $(R_r\cos\varphi_m, R_r\sin\varphi_m, 0)$ in $\bar{Z}\!\!-\!\!\bar{X}\bar{O}\bar{Y}$ coordinate system. After coordinate transformation, the coordinate of the $m$th antenna element on the transmit UCA in $Z'\!\!-\!\!X'\bar{O}Y'$ coordinate system, denoted by $(B'_x,B'_y,B'_z)$, can be expressed as
\begin{equation} \label{B1}
\begin{bmatrix}B'_x \\ B'_y \\ B'_z\end{bmatrix}=
\mathbf{R}_y(\gamma)\mathbf{R}_x(\alpha)
\begin{bmatrix}R_r\cos\varphi_m \\ R_r\sin\varphi_m \\ 0 \end{bmatrix},
\end{equation}
where $\mathbf{R}_x(\alpha)$ and $\mathbf{R}_y(\gamma)$ represent the rotation matrices corresponding to the x-axis and y-axis, respectively, which are given by
\begin{equation} \label{Rx}
\mathbf{R}_x(\alpha)=
\begin{bmatrix}1 & 0 & 0 \\ 0 & \cos\alpha & -\sin\alpha \\ 0 & \sin\alpha & \cos\alpha \end{bmatrix},
\end{equation}
\begin{equation} \label{Ry}
\mathbf{R}_y(\gamma)=
\begin{bmatrix} \cos\gamma & 0 & \sin\gamma\\ 0 & 1 & 0 \\ -\sin\gamma & 0 & \cos\gamma  \end{bmatrix}.
\end{equation}
The coordinate system $Z'\!\!-\!\!X'\bar{O}Y'$ is parallel to $Z\!\!-\!\!XOY$, and thus the coordinate of the $m$th antenna element on the receive UCA in $Z\!\!-\!\!XOY$ coordinate system can be expressed as
\begin{equation} \label{B}
\begin{bmatrix}B_x \\ B_y \\ B_z\end{bmatrix}=
\begin{bmatrix}B'_x \\ B'_y \\ B'_z\end{bmatrix}+
\begin{bmatrix}0 \\ 0 \\ D \end{bmatrix}.
\end{equation}
After matrix multiplication and addition, the coordinate of the $m$th antenna element on the receive UCA in $Z\!\!-\!\!XOY$ coordinate system is shown in \eqref{Bxyz}.



\bibliographystyle{IEEEtran}
\bibliography{OAM}

\begin{thebibliography}{10}
\providecommand{\url}[1]{#1}
\csname url@samestyle\endcsname
\providecommand{\newblock}{\relax}
\providecommand{\bibinfo}[2]{#2}
\providecommand{\BIBentrySTDinterwordspacing}{\spaceskip=0pt\relax}
\providecommand{\BIBentryALTinterwordstretchfactor}{4}
\providecommand{\BIBentryALTinterwordspacing}{\spaceskip=\fontdimen2\font plus
\BIBentryALTinterwordstretchfactor\fontdimen3\font minus
  \fontdimen4\font\relax}
\providecommand{\BIBforeignlanguage}[2]{{%
\expandafter\ifx\csname l@#1\endcsname\relax
\typeout{** WARNING: IEEEtran.bst: No hyphenation pattern has been}%
\typeout{** loaded for the language `#1'. Using the pattern for}%
\typeout{** the default language instead.}%
\else
\language=\csname l@#1\endcsname
\fi
#2}}
\providecommand{\BIBdecl}{\relax}
\BIBdecl

\bibitem{WRC}
(2018, Sep.) World radiocommunication conference ({WRC}). [Online]. Available:
  \url{https://www.itu.int/en/ITU-R/conferences/wrc/ Pages/default.aspx}.

\bibitem{Song2011Present}
H.~{Song} and T.~{Nagatsuma}, ``Present and future of terahertz
  communications,'' \emph{IEEE Trans. Terahertz Sci. Technol.}, vol.~1, no.~1,
  pp. 256--263, 2011.

\bibitem{Ghosh20195G}
A.~{Ghosh}, A.~{Maeder}, M.~{Baker}, and D.~{Chandramouli}, ``5{G} evolution: A
  view on 5{G} cellular technology beyond 3{GPP} release 15,'' \emph{IEEE
  Access}, vol.~7, pp. 127\,639--127\,651, 2019.

\bibitem{Popovski5G}
P.~{Popovski}, K.~F. {Trillingsgaard}, O.~{Simeone}, and G.~{Durisi}, ``5{G}
  wireless network slicing for e{MBB}, {URLLC}, and m{MTC}: A
  communication-theoretic view,'' \emph{IEEE Access}, vol.~6, pp.
  55\,765--55\,779, 2018.

\bibitem{3GPP2017}
3GPP, ``{TS} 38.101-1 v0.0.1; {NR}; user equipment ({UE}) radio transmission
  and reception,'' {3rd Generation Partnership Project (3GPP)}, Tech. Rep.,
  2017.

\bibitem{Allen1992Orbital}
L.~Allen, M.~W. Beijersbergen, R.~J. Spreeuw, and J.~P. Woerdman, ``Orbital
  angular momentum of light and the transformation of {L}aguerre-{G}aussian
  laser modes,'' \emph{Phys. Rev. A: At. Mol. Opt. Phys.}, vol.~45, no.~11, pp.
  8185--8189, 1992.

\bibitem{Tamburini2012Encoding}
F.~Tamburini, E.~Mari, A.~Sponselli, F.~Romanato, T.~Bo, A.~Bianchini,
  L.~Palmieri, and G.~Someda, ``Encoding many channels on the same frequency
  through radio vorticity: first experimental test,'' \emph{New J. Phys.},
  vol.~14, no.~3, p. 033001, Mar. 2012.

\bibitem{Mahmouli20134}
F.~E. Mahmouli and S.~D. Walker, ``4-{G}bps uncompressed video transmission
  over a 60-{GH}z orbital angular momentum wireless channel,'' \emph{IEEE
  Wireless Commun. Lett.}, vol.~2, no.~2, pp. 223--226, Apr. 2013.

\bibitem{Yan2014High}
Y.~Yan, G.~Xie, and et~al., ``High-capacity millimetre-wave communications with
  orbital angular momentum multiplexing,'' \emph{Nat. Commun.}, vol.~5, p.
  4876, 09 2014.

\bibitem{Zhang2016The}
Z.~Zhang, S.~Zheng, Y.~Chen, X.~Jin, H.~Chi, and X.~Zhang, ``The capacity gain
  of orbital angular momentum based multiple-input-multiple-output system.''
  \emph{Sci. Rep.}, vol.~6, no.~1, pp. 25\,418--25\,418, May 2016.

\bibitem{Ren2017Line}
Y.~{Ren}, L.~{Li}, and et~al., ``Line-of-sight millimeter-wave communications
  using orbital angular momentum multiplexing combined with conventional
  spatial multiplexing,'' \emph{IEEE Trans. Wireless Commun.}, vol.~16, no.~5,
  pp. 3151--3161, May 2017.

\bibitem{Zhang2017Mode}
W.~Zhang, S.~Zheng, X.~Hui, R.~Dong, X.~Jin, H.~Chi, and X.~Zhang, ``Mode
  division multiplexing communication using microwave orbital angular momentum:
  An experimental study,'' \emph{IEEE Trans. Wireless Commun.}, vol.~16, no.~2,
  pp. 1308--1318, Feb. 2017.

\bibitem{Chen2018A}
R.~Chen, W.~Yang, H.~Xu, and J.~Li, ``A {2-D FFT}-based transceiver
  architecture for {OAM-OFDM} systems with {UCA} antennas,'' \emph{IEEE Trans.
  Veh. Commun.}, vol.~67, no.~6, pp. 5481--5485, Jun. 2018.

\bibitem{Chen2018Beam}
R.~Chen, H.~Xu, M.~Moretti, and J.~Li, ``Beam steering for the misalignment in
  {UCA}-based {OAM} communication systems,'' \emph{IEEE Wireless Commun.
  Lett.}, vol.~7, no.~4, pp. 582--585, Aug. 2018.

\bibitem{Chen2019On}
R.~{Chen}, H.~{Xu}, X.~{Wang}, and J.~{Li}, ``On the performance of {OAM} in
  keyhole channels,'' \emph{IEEE Wireless Commun. Lett.}, vol.~8, no.~1, pp.
  313--316, Feb. 2019.

\bibitem{Zhang2019Orbital}
C.~{Zhang} and Y.~{Zhao}, ``Orbital angular momentum nondegenerate index
  mapping for long distance transmission,'' \emph{IEEE Trans. Wireless
  Commun.}, vol.~18, no.~11, pp. 5027--5036, Nov. 2019.

\bibitem{Chen2019Orbital}
R.~{Chen}, H.~{Zhou}, M.~{Moretti}, X.~{Wang}, and J.~{Li}, ``Orbital angular
  momentum waves: Generation, detection and emerging applications,'' \emph{IEEE
  Commun. Surv. Tut.}, vol.~99, no.~1, pp. 1--30, Nov. 2019.

\bibitem{Zhao2019Compound}
Y.~{Zhao} and C.~{Zhang}, ``Compound angular lens for radio orbital angular
  momentum coaxial separation and convergence,'' \emph{IEEE Antennas Wireless
  Propag. Lett.}, vol.~18, no.~10, pp. 2160--2164, Oct. 2019.

\bibitem{Chen2020OAM}
R.~{Chen}, Z.~{Tian}, H.~{Zhou}, and W.~{Long}, ``{OAM}-based concentric
  spatial division multiplexing for cellular {IoT} terminals,'' \emph{IEEE
  Access}, vol.~8, pp. 59\,659--59\,669, 2020.

\bibitem{Chen2020Multi}
R.~{Chen}, W.~X. {Long}, X.~{Wang}, and J.~{Li}, ``Multi-mode {OAM} radio
  waves: Generation, angle of arrival estimation and reception with {UCAs},''
  \emph{IEEE Trans. Wireless Commun.}, vol.~19, no.~10, pp. 6932--6947, 2020.

\bibitem{Chen2020Generation}
R.~{Chen}, M.~{Zou}, X.~{Wang}, and A.~{Tennant}, ``Generation and beam
  steering of arbitrary-order {OAM} with time-modulated circular arrays,''
  \emph{IEEE Systems Journal}, pp. 1--8, 2020.

\bibitem{Edfors2012Is}
O.~Edfors and A.~J. Johansson, ``Is orbital angular momentum ({OAM}) based
  radio communication an unexploited area?'' \emph{IEEE Trans. Antennas
  Propag.}, vol.~60, no.~2, pp. 1126--1131, Feb. 2012.

\bibitem{Zhang20196G}
Z.~{Zhang}, Y.~{Xiao}, Z.~{Ma}, M.~{Xiao}, Z.~{Ding}, X.~{Lei}, G.~K.
  {Karagiannidis}, and P.~{Fan}, ``6{G} wireless networks: Vision,
  requirements, architecture, and key technologies,'' \emph{IEEE Veh. Technol.
  Mag.}, vol.~14, no.~3, pp. 28--41, Sep. 2019.

\bibitem{Yang20196G}
P.~{Yang}, Y.~{Xiao}, M.~{Xiao}, and S.~{Li}, ``6{G} wireless communications:
  Vision and potential techniques,'' \emph{IEEE Network}, vol.~33, no.~4, pp.
  70--75, Jul. 2019.

\bibitem{YjZhang2013}
Y.~Zhang, W.~Feng, and N.~Ge, ``On the restriction of utilizing orbital angular
  momentum in radio communications,'' in \emph{8th Int. Conf. Commun.
  Networking China}, Aug. 2013, pp. 271--275.

\bibitem{Xie2015Performance}
G.~Xie, L.~Li, Y.~Ren, and et~al., ``Performance metrics and design
  considerations for a free-space optical orbital-angular-momentum-multiplexed
  communication link,'' \emph{Optica}, vol.~2, no.~4, pp. 357--365, 2015.

\bibitem{Zaidi2018OFDM}
A.~A. {Zaidi}, R.~{Baldemair}, V.~{Moles-Cases}, N.~{He}, K.~{Werner}, and
  A.~{Cedergren}, ``{OFDM} numerology design for {5G} new radio to support
  {IoT}, {eMBB}, and {MBSFN},'' \emph{IEEE Commun. Stand. Mag.}, vol.~2, no.~2,
  pp. 78--83, Jun. 2018.

\bibitem{Mohammadi2010Orbital}
S.~M. Mohammadi, L.~K.~S. Daldorff, J.~E.~S. Bergman, R.~L. Karlsson,
  B.~Thid\'e, K.~Forozesh, T.~D. Carozzi, and B.~Isham, ``Orbital angular
  momentum in radio¡ª{A} system study,'' \emph{IEEE Trans. Antennas Propag.},
  vol.~58, no.~2, pp. 565--572, Feb. 2010.

\bibitem{Marzetta2010Noncooperative}
T.~L. {Marzetta}, ``Noncooperative cellular wireless with unlimited numbers of
  base station antennas,'' vol.~9, no.~11, pp. 3590--3600, 2010.

\bibitem{Chen2017Misalignment}
R.~{Chen}, H.~{Xu}, J.~{Li}, and Y.~{Zhang}, ``Misalignment-robust receiving
  scheme for {UCA}-based {OAM} communication systems,'' in \emph{IEEE 85th Veh.
  Technol. Conf.}, June 2017, pp. 1--5.

\end{thebibliography}
\end{document}